\documentclass[10pt, journal]{IEEEtran}

\RequirePackage{rotating}

\ifCLASSINFOpdf
\else
  \usepackage[dvips]{graphicx}
\fi
\usepackage{url}

\usepackage{graphicx}
\usepackage{cite}
\usepackage{booktabs}
\usepackage{tabularx}
\usepackage{float}
\usepackage{balance}
\usepackage{flushend}
\usepackage{amssymb, amsmath, amsfonts}
\usepackage{array}
\usepackage{authblk}

\DeclareGraphicsExtensions{.png}

\newcommand*{\Z}{\mathbb{Z}}

\begin{document}

\title{Longitudinal Acoustic Speech Tracking Following Pediatric Traumatic Brain Injury}
\author[1, 2]{Camille Noufi}
\author[2,3]{Adam C. Lammert}
\author[2,4]{Daryush D. Mehta}
\author[2]{James R. Williamson}
\author[2]{Gregory Ciccarelli}
\author[2]{Douglas Sturim}
\author[2,4]{Jordan R. Green}
\author[2]{Thomas F. Quatieri}
\author[5]{Thomas F. Campbell
\thanks{Approved for public release. Distribution is unlimited. This material is based upon work supported by the Under Secretary of Defense for Research and Engineering under Air Force Contract No. FA8702-15-D-0001. Any opinions, findings, conclusions or recommendations expressed in this material are those of the author(s) and do not necessarily reflect the views of the Under Secretary of Defense for Research and Engineering. This research was supported, in part, by the National Institute on Deafness and other Communication Disorders Grant R01DC0368. \\Correspondence may be sent to C. Noufi (e-mail: cnoufi@ccrma.stanford.edu).}
}
\affil[1]{Stanford University, Stanford, CA, USA}
\affil[2]{MIT Lincoln Laboratory, Lexington, MA, USA}
\affil[3]{Worcester Polytechnic Institute, Worcester, MA, USA}
\affil[4]{Mass General Hospital, Boston, MA, USA}
\affil[5]{University of Texas at Dallas, Dallas, TX, USA}

\maketitle

\begin{abstract}
Recommendations for common outcome measures following pediatric traumatic brain injury (TBI) support the integration of instrumental measurements alongside perceptual assessment in recovery and treatment plans. A comprehensive set of sensitive, robust and non-invasive measurements is therefore essential in assessing variations in speech characteristics over time following pediatric TBI. In this article, we study the changes in the acoustic speech patterns of a pediatric cohort of ten subjects diagnosed with severe TBI. We extract a diverse set of both well-known and novel acoustic features from child speech recorded throughout the year after the child produced intelligible words. These features are analyzed individually and by speech subsystem, within-subject and across the cohort. As a group, older children exhibit highly significant ($p<0.01$) increases in pitch variation and phoneme diversity, shortened pause length, and steadying articulation rate variability. Younger children exhibit similar steadied rate variability alongside an increase in formant-based articulation complexity. Correlation analysis of the feature set with age and comparisons to normative developmental data confirm that age at injury plays a significant role in framing the recovery trajectory. Nearly all speech features significantly change ($p<0.05$) for the cohort as a whole, confirming that acoustic measures supplementing perceptual assessment are needed to identify efficacious treatment targets for speech therapy following TBI.
\end{abstract}


\section{Introduction}

Speech generation calls upon a large and distributed neural network in the brain, and is a good measure of the state of the brain following a head injury. Impairment to the brain's widely distributed speech network commonly occurs following traumatic brain injury (TBI) and often manifests in a voice or speech disorder, persistent dysarthria being the most frequently reported voice disorder in school-aged children and adolescents after sustaining a TBI~\cite{murdoch-artic-tbi}. The effects of TBI present themselves within variants of this disorder as well. Darley, Aronson and Brown defined dysarthria as `a collective name for a group of speech disorders resulting from disturbances in muscular control over the speech mechanism due to damage of the central peripheral nervous system~\cite{dysarthria-def}. It may be specifically classified as spastic, hyper/hypokinetic, flaccid, ataxic or mixed, with spastic and mixed being highlighted as the most prevalent. Murdoch \textit{et al.} provide a detailed description of the neurological bases of these subtypes~\cite{Murdoch-neuro} and their associated clinical features~\cite{Zafonte} that often manifest within the articulatory, prosodic and phonatory speech subsystems~\cite{Murdoch, Zafonte, dysarthria_acoustic, book-acoustic-dysarthria}. A perceptual study by Murdoch \textit{et al.} of 22 children who sustained a TBI found significant disturbance in various aspects of prosody, resonance, phonation and articulation, and classified 10 of the 22 subjects as dysarthric~\cite{Murdoch}. In cases of amyotrophic lateral sclerosis (ALS), Rong \textit{et al.} found that features within the articulatory and phonatory subsystems provided more responsive cues to early stages of bulbar deterioration than standard assessments of intelligibility and speaking rate~\cite{rong2015}.  A close look at these affected speech subsystems can provide refined cues to impairments of the brain's speech network as well as identify efficacious treatment targets for speech therapy~\cite{book-acoustic-dysarthria, rong2015}.

Recommendations for common outcome measures following pediatric TBI support the integration of objective, instrumental measurement alongside perceptual assessment in recovery and treatment plans~\cite{McCauley-RecsTBI, book-acoustic-dysarthria}. Physiological assessment tools are commonly used to measure physical attributes of impairment~\cite{Murdoch,dysarthria_acoustic, murdoch-artic-tbi} and acoustic features used to identify impaired speech have been studied for several decades~\cite{book-acoustic-dysarthria}. Many acoustic analysis programs are commercially available~\cite{Zafonte}. More recently, case-control studies of adult populations have used audio-based feature extraction techniques to classify cognitive state following head injury~\cite{mTBI,vowels_mtbi,mtbi-concussion,mbti-lammert}. Related studies have used acoustic features to examine similar types of speech degradation due to ALS~\cite{book-acoustic-dysarthria, Green-ALS, rong2015}, depression~\cite{avec_paper,depression_2013,helfter_depression_2013,phone_class}, Parkinson's disease~\cite{book-acoustic-dysarthria}, cognitive load~\cite{book-quatieri} and diagnosed dysarthria and dysphonia~\cite{book-acoustic-dysarthria,cpp,cpp-predict,breathy_acoustics}. 

Despite the need for objective and repeatable measurements in assessing TBI-affected pediatric speech, few acoustic-based or longitudinal studies have been performed on this population~\cite{Murdoch,Zafonte}. Longitudinal, signal-based analysis of speech following pediatric TBI has thus far only examined speaking rate~\cite{campbell-rate}. Campbell \textit{et al.} demonstrated that syllabic speaking rate was initially much lower in TBI-affected children than in the control group, but rate increase among subjects was seen over time. Articulatory rate versus pause time was examined and the authors hypothesized that each were independent contributors to overall slowed speaking rate and separate windows into linguistic processing. Initial speaking rates and recovery speeds varied across the group with TBI but these variations were not studied further. However, additional perceptual longitudinal studies by Campbell \textit{et al.} demonstrated that although several motor-speech characteristics are commonly impaired following TBI, rate and trajectory of recovery varies and is likely age-dependent. A study focused on consonant accuracy found that children under 5 years old were less likely to reach the normal accuracy range for their age following TBI than older children were~\cite{PCC-R}. Another study was the first to focus on the longitudinal change in expressive speech during spontaneous conversation following pediatric TBI~\cite{campbell_numwords_long}. It noted the importance of including concomitant factors such as age in the analysis. A developmental study of healthy children by Nip and Green found that increases in speaking rate with age were heavily due to gains in cognitive and linguistic processing and speech motor control, and that these gains primarily took place between 4 and 13 years old~~\cite{green-processing-rate2013}. Moreover, Lee \textit{et al.} used acoustic phoneme measurements to track normative speech-motor development trajectories in children~\cite{Lee-Acoustics}. 

Thus, in this study we examine TBI-affected speech in a pediatric population over time by pooling together a diverse set of acoustic measurements spanning three speech subsystems. These features are measured monthly over the course of one year to track how particular acoustic characteristics of each speech subsystem are changing in relation to each other. Section 2 details this methodology, originally presented by the authors in~\cite{childtbi}. In Section 3, we expand upon our analysis of the cohort-wide trends, provide new analyses on nuanced relationship between recovery and age, and present case-studies of individual subjects. Section 4 suggests future research based on our findings and Section 5 concludes our work. 

\section{Methods}

\subsection{Study Design and Data Collection}

We apply our feature analysis to speech data originally collected by Campbell \textit{et al.} to understand longitudinal changes in the perceptual `Percent Consonants Correct - Revised' (PCC-R) intelligibility measure. This data consists of recorded conversations between a subject and a trained examiner~\cite{PCC-R}. Due to the large amount of preprocessing required to utilize this dataset for acoustic analysis, we focus on ten out of fifty-six subjects within the original study. The ten-subject cohort was selected to be representative of the age range within the original sample, include both males and females, and have corresponding recordings of good perceptual quality. 

\begin{figure}[t]
 \centering
 \includegraphics[width=\linewidth]{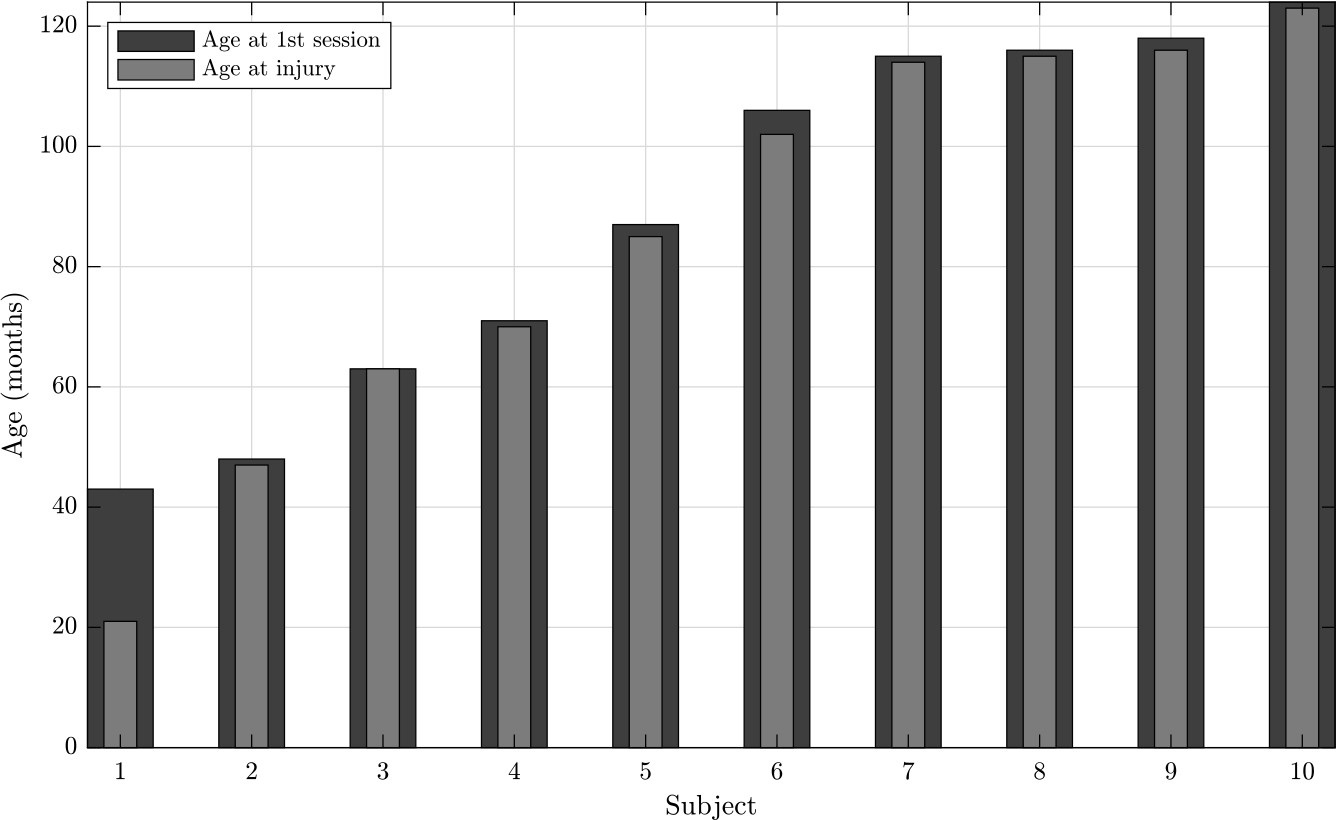}
 \caption{Distribution of subjects by age (in months).}
 \label{fig:cohort_age}
\end{figure}

Each subject had been recently diagnosed with a severe TBI. Recorded speech sampling sessions began as soon as possible after the child was able to produce at least 10 intelligible words. Available metadata for each subject consisted of gender, age in months at injury, and age in months at the first speech sampling session. The cohort consisted of six males and four females ranging from 43 months to 123 months in age at the first session. Fig.~\ref{fig:cohort_age} shows the distribution of subjects by age. 

Twelve speech sampling sessions were scheduled to occur monthly; we were able to obtain $(\mu,\sigma) = (11.5,0.7)$ sessions per subject. The collection period occurred between 1999 and 2005. All speech samples were recorded at a sampling rate of 44.1kHz using a Shure UHF wireless or Radio Shack 33-3003 microphone and a Marantz PMD 402/502 audio cassette recorder. Campbell \textit{et al.} provide a thorough description of participant eligibility, original study design and data collection in~\cite{PCC-R}.

\subsection{Data Preprocessing}

Several preprocessing steps were necessary to measure signal-based changes to a child's speech within and across sessions. The raw audio consisted of 15 to 40 minutes of conversational speech between two to four speakers (occasionally family members were present during a session). Because the speech was originally obtained for perceptual assessment, steps to reduce channel effects and noise were not needed during the recording process. Therefore, speaker segmentation and channel-noise reduction were applied to all recordings.

Speaker separation was performed via a Gaussian-mixture-model-based diarization tool using 100 ms minimum duration segments and 2 speaker clusters~\cite{gmm1,gmm2}. However, diarization models are rarely trained on child speech, so manual tuning was performed before and after automatic diarization to ensure the maximum amount of a subject's speech was extracted while excluding other speakers, lengthy silences and non-stationary noise. This was done by selecting the timestamps of child speech detected by the diarization tool, padding each segment with an additional 0.2 seconds to retain lengthy consonant utterances, and then fusing together segments less than 1 second apart. Segments less than 2 seconds in length were discarded after this step, as they consisted of fragmented speech that would confound prosody and timing measurements in downstream processing. Each file was then manually checked for any remaining cross-talk or other perceivable non-child-speech elements. The intact child speech was enhanced using the Optimal Modified Minimum Mean-Square Error Log-Spectral Amplitude method~\cite{OMSLA} to remove additive non-stationary noise while limiting distortion of clean speech. We were able to extract $(\mu,\sigma) = (2.14,1.23)$ minutes of clean, connected child speech from each session recording. 

\subsection{Feature Extraction}

\subsubsection{Formant-Track Coordination}

Many studies report that TBI may degrade the physiological coordination underlying vocal tract trajectories and reduce the magnitude and complexity of articulator movement during speech~\cite{Murdoch,Zafonte,book-acoustic-dysarthria,mTBI,mtbi-concussion,mtbi-portable}. Vocal tract resonance trajectories are captured in the time series of audio-based formant frequencies. We use multivariate auto- and cross-correlations as a proxy measure of coordination between these formant-tracks. This high-level feature is designed to capture dynamical interrelations among the multiple articulators involved in speech production. It is motivated by the observation that auto- and cross-correlations of measured signals can reveal hidden parameters in the stochastic-dynamical systems that generate the time series. Changes over time in the coupling strengths among the formant tracks cause changes in the eigenvalue spectra of the resulting correlation matrices; weakly coupled formant-tracks may indicate more complex interactions between the articulators. Williamson \textit{et al.} first applied this multivariate correlation approach to epileptic seizure prediction from multichannel EEG~\cite{williamson_eeg} and subsequently to the tracking and prediction of major depressive disorder from audio-based vocal signals~\cite{avec_paper,Williamson-xcorr-tracking2019}. 

We tailor this implementation to fit our session recordings of variable length and content. The first three formant-tracks (F1-F3) are estimated using the extended Kalman algorithm by Mehta and Rudoy~\cite{KARMA}. The algorithm uses a 10 ms frame interval and is seeded with average age- and gender-dependent formant values reported by Lee \textit{et al.}~\cite{Lee-Acoustics} . Embedded in the formant tracking algorithm is an energy-based voice-activity detector that allows a Kalman smoother to coast through non-speech regions. Estimates of the F3 above a threshold of 4.5k Hz are truncated. The tracks are then subdivided into 20 second segments; segments at the end of the recording are discarded if under 5 seconds. 

For each 20 second segment, a channel-delay correlation matrix is computed from the three low-level formant tracks. Each matrix contains correlation coefficients between the tracks at a relative time delay. Five matrices are computed at five delay scales (10, 30, 70, 100 and 150 ms) with 15 time-delays used per scale. The eigenvalues of a matrix are extracted by rank-order. The resulting eigenvalue spectrum represents the sizes of the principal axes of the time-embedded scatter distribution embedded in each correlation matrix. Williamson \textit{et al.} provide a complete mathematical description of this method in~\cite{Williamson-xcorr-tracking2019}.

For each delay scale, the resulting eigenvalue spectra of the 20 second segments are averaged to obtain the mean eigenvalue decomposition for the session. This approach is performed per session, per subject. 

\begin{table}[tbhp]
\centering
\caption{Acoustic measurements and their related speech characteristic (CO=Consonant, VW=Vowel)}
\label{tab:phone_measures}
\begin{tabular}{@{}ll@{}}
\toprule
\textbf{Measurement} & \textbf{Speech Characteristic} \\ \midrule
F1-F3 Auto- and Cross-Correlation & Vocal Tract Coordination \\
Speech vs. Articulation Rate & Intraphrase Pauses \\
CO and VW Duration & Articulation Rate \\
Variance of CO and VW Duration & Rate variability \\
\% Phonemes are CO & Complexity and Precision \\
Number of Unique Phonemes & Complexity and Precision \\
F0 Mean & Abnormal Pitch \\
F0 Variance & Monotonous Pitch \\
CPP & Breathiness/Dysphonic Quality \\
H2-H1 & Creakiness \\
\bottomrule
\end{tabular}
\end{table}

\subsubsection{Phoneme-based Prosody and Timing}

We measure prosody at the phoneme level under the hypothesis that this captures nuanced information in motor and linguistic planning, timing, and execution~\cite{book-acoustic-dysarthria,dysarthria_acoustic,vowels_mtbi,phone_class}. The KALDI ASpIRE neural network model is used to extract all uttered phonemes and silences from a session recording and tag them with predicted English phoneme type and duration~\cite{kaldi}. We perform a statistical summary across the phoneme collection to obtain the mean, median, variance, inter-quartile range, and count of all 39 Standard American English phonemes and silence regions (assumed as pauses) per session. We combine these statistics to create the rate and complexity measurements in Table~\ref{tab:phone_measures}. Speaking rate includes pause duration whereas articulation rate measures phoneme duration only. 

\subsubsection{Phonation}
Cepstral Peak Prominence (CPP) and the H2-H1 relative amplitude difference (H2H1) have been shown to quantify breathy and dysphonic~\cite{cpp,cpp2,cpp3,cpp-predict,cpp-index,cpp-cutoff,breathy_acoustics}, and creaky phonation~\cite{breathy_acoustics,h2h1}, respectively. Breathiness displays low periodicity and steep spectral slope in the source spectrum; its cepstral representation contains peaks with smaller amplitudes. CPP measures the difference between these peaks and the surrounding noise floor.  A lower CPP is therefore an indication of a more breathy voice. The source spectrum of a creaky voice due to glottal constriction often contains a flatter spectral tilt and a much larger second harmonic (H2) amplitude than first harmonic (H1) amplitude. A large H2H1 therefore indicates a higher probability of creaky phonation due to glottal constriction.

To measure phonatory abnormalities, we first utilize the Praat software tool\footnote{http://www.fon.hum.uva.nl/praat/} to extract fundamental frequency (F0) estimations and voice-activity regions via autocorrelation~\cite{pitchtrack}. We measure CPP and H2H1 at 10 ms frame intervals (using a 20 ms window size with an overlap of 50\%) and mask out frames labeled as voiced. We then perform a statistical summary within each session to obtain the mean, median, skew, variance, and inter-quartile range as our quantitative representation of breathy and creaky voice quality for the session. 

\subsection{Trend Summarization}
\label{method_sum}

Longitudinal changes to these features are measured via linear correlation with time.  These correlations summarize the direction and strength of change to a speech feature between the initial and final sampling sessions.     

Formant-track coordination complexity is summarized over time by examining how the eigenvalue spectrum of a correlation matrix changes over the course of 12 months. The average eigenvalue spectrum of each session is linearly correlated per-rank with session number (1 is the first session, 12 is the final session, approximately 12 months later) to obtain a vector describing this change over time. Equation~\ref{eq_fcorr} defines this vector for a single subject at one delay scale.

\begin{equation}
 (\rho_\lambda[r],p_\lambda[r]) = corr(\lambda_r,S) \; \forall r \in \Z ^{1:(M \times N)}
 \label{eq_fcorr}
\end{equation}

\noindent where $\lambda_r$ is the eigenvalue of a particular session at rank $r$, $M$ is the number of channels, $N$ is the number of delays per channel, $S\in \Z ^{1 \times i}$ is the vector containing all sessions, and $i$ is the number of sessions. \textbf{$\rho_\lambda$} is the rank-ordered vector describing the formant-track correlation with and $\bf{p}$ is the p-value significance vector.

\begin{figure}[thbp]
 \centering
 \includegraphics[width=\linewidth]{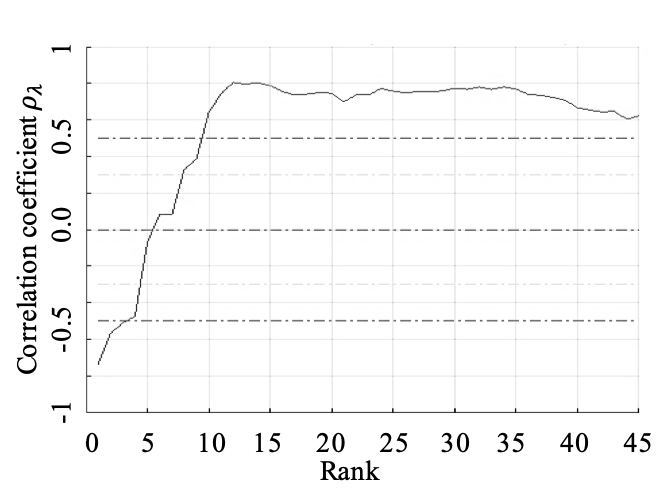}
 \caption{Correlation coefficients $\bf{\rho_\lambda}$ of session number and F1-F3 correlation matrix eigenvalues at a 70 ms delay scale. This delay captures the coupling strength between the formant tracks at a wider time interval, capturing the presence of any coordination delay.  This time-based correlation is performed for all delay scales.  The 70 ms case is shown for brevity.}
 \label{fig:formant_eig_corr}
\end{figure}

Fig.~\ref{fig:formant_eig_corr} displays the correlation $\bf{\rho_\lambda}$ of the youngest subject's formant-track interrelations with time at a 70 ms delay scale.  Here, low-rank eigenvalues negatively correlate with time and higher-rank eigenvalues positively correlate with time. This represents information distributing throughout the correlation matrix and the time-embedded formant scatter distribution becoming more `complex' throughout the 12 sessions.

Phoneme-based and phonatory changes over time are summarized via linear correlation of the summary statistics with session number. An individual subject's summary statistics are correlated with his/her specific $S$ vector (see~\ref{eq_fcorr}). Age-independent group trends are calculated by first normalizing each subject's measurements across all sessions to zero mean and unit standard deviation (e.g. $zscore(CPP_{mean})$) to remove effects of age-influenced baseline values. These normalized measurements are then concatenated. $S$ vectors for each subject are concatenated as well. For each speech characteristic $c_n$, a Pearson correlation is performed between these two aggregate vectors to obtain correlation coefficient $\rho_n$ and significance p-value $p_n$. These age-independent group trends are calculated for all ten subjects, for younger children (less than 72 months old), and for older children (72 months and older). To capture developmental influence on baseline measurements and trends, age-dependent group correlations are calculated by replacing each subject's $S$ vector with his/her age (in months) at each session and skipping normalization.

\begin{table}[bthp]
\centering
\caption{Trend profile of cohort using Pearson correlation to indicate direction and strength of speech changes over one year. YP = younger children, OP = older children. Bold values indicate measures of high significance with $p \leq 0.01$.}
\label{tab:atp_group}
\renewcommand{\tabcolsep}{1mm}
\begin{tabular}{@{}ll|cccc@{}}
\toprule
\textbf{Articulation} & \textbf{} & \textbf{YP} & \textbf{OP} & \textbf{All} & \textbf{With Age} \\ \midrule
Formant Coordination & & \textbf{$\rho$ = 0.50} & 0.05 & 0.22 & 0.05 \\
 & & \textbf{p = 0.01} & \textgreater 0.1 & \textgreater 0.1 & 0.6 \\ \midrule
Consonant/ & Duration & -0.052 & 0.013 & 0.043 & 0.547 \\
Vowel Ratio & & 0.728 & 0.917 & 0.65 & 0 \\
 & Occurrence & 0.253 & 0.167 & 0.193 & 0.404 \\
 & & 0.086 & 0.174 & 0.039 & 0 \\ \midrule
\# Unique Phonemes & & 0.241 & \textbf{0.438} & \textbf{0.357} & 0.322 \\
 & & 0.102 & \textbf{0} & \textbf{0} & 0.103 \\ \midrule
\textbf{Prosody} & \textbf{} & \textbf{YP} & \textbf{OP} & \textbf{All} & \textbf{With Age} \\ \midrule
Overall Rate & Speech & 0.329 & \textbf{0.513} & \textbf{0.437} & 0.568 \\
 & & 0.024 & \textbf{0} & \textbf{0} & 0 \\
 & Articulation & 0.196 & 0.202 & 0.199 & 0.618 \\
 & & 0.188 & 0.099 & 0.033 & 0 \\ \midrule
Consonant & Duration & -0.261 & -0.225 & \textbf{-0.240} & -0.495 \\
 & & 0.076 & 0.065 & \textbf{0.010} & 0 \\
 & Variation & \textbf{-0.505} & \textbf{-0.390} & \textbf{-0.437} & -0.311 \\
 & & \textbf{0} & \textbf{0.001} & \textbf{0} & 0.001 \\ \midrule
Vowel & Duration & -0.183 & -0.213 & -0.201 & -0.604 \\
 & & 0.218 & 0.081 & 0.032 & 0 \\
 & Variation & -0.318 & -0.259 & \textbf{-0.283} & -0.42 \\
 & & 0.029 & 0.033 & \textbf{0.002} & 0 \\ \midrule
Pitch & Mean & 0.325 & \textbf{-0.692} & -0.325 & -0.776 \\
 & & 0.026 & \textbf{0} & 0.026 & 0 \\
 & Variation & -0.301 & \textbf{0.538} & -0.301 & -0.688 \\
 & & 0.04 & \textbf{0} & 0.039 & 0 \\ \midrule
\textbf{Phonation} & \textbf{} & \textbf{YP} & \textbf{OP} & \textbf{All} & \textbf{With Age} \\ \midrule
Breathy Voice & Mean & -0.355 & \textbf{-0.317} & \textbf{-0.306} & 0.455 \\
 & & 0.014 & \textbf{0.009} & \textbf{0.001} & 0 \\
 & Variation & 0.207 & 0.232 & 0.222 & -0.395 \\
 & & 0.163 & 0.057 & 0.017 & 0 \\ \midrule
Creaky Voice & Mean & \textbf{0.471} & \textbf{0.469} & 0.102 & 0.423 \\
 & & \textbf{0.001} & \textbf{0} & 0.271 & 0 \\
 & Skew & \textbf{-0.418} & \textbf{-0.319} & -0.015 & -0.315 \\
 & & \textbf{0.003} & \textbf{0.008} & 0.871 & 0.001 \\ \bottomrule
\end{tabular}
\end{table}

We group this set of features by speech subsystem and produce a trend profile as a comprehensive summarization. Change in formant-track complexity is represented by the correlation of the first-order eigenvalue with session number $\rho_\lambda[1]$. Correlation coefficients and significance values of mean and variance summarize the change to phoneme-based and phonation features.  

\section{Results and Discussion}

\subsection{Aggregate Trends}

\begin{figure*}[hbtp]
 \centering
 \includegraphics[width=\linewidth]{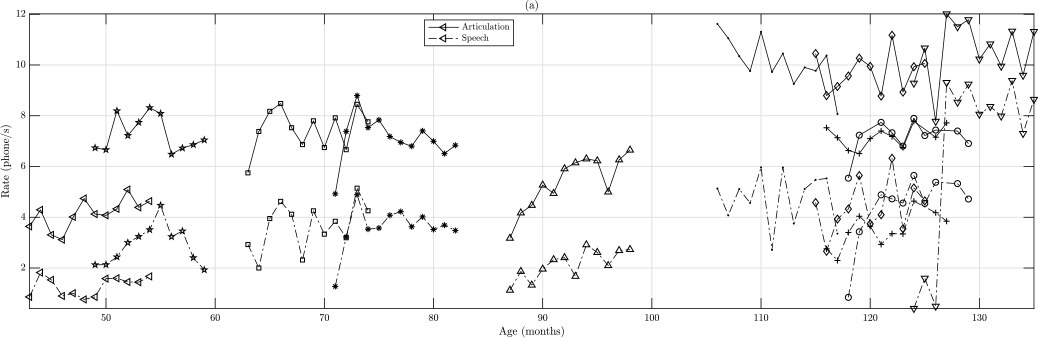}
 \includegraphics[width=\linewidth]{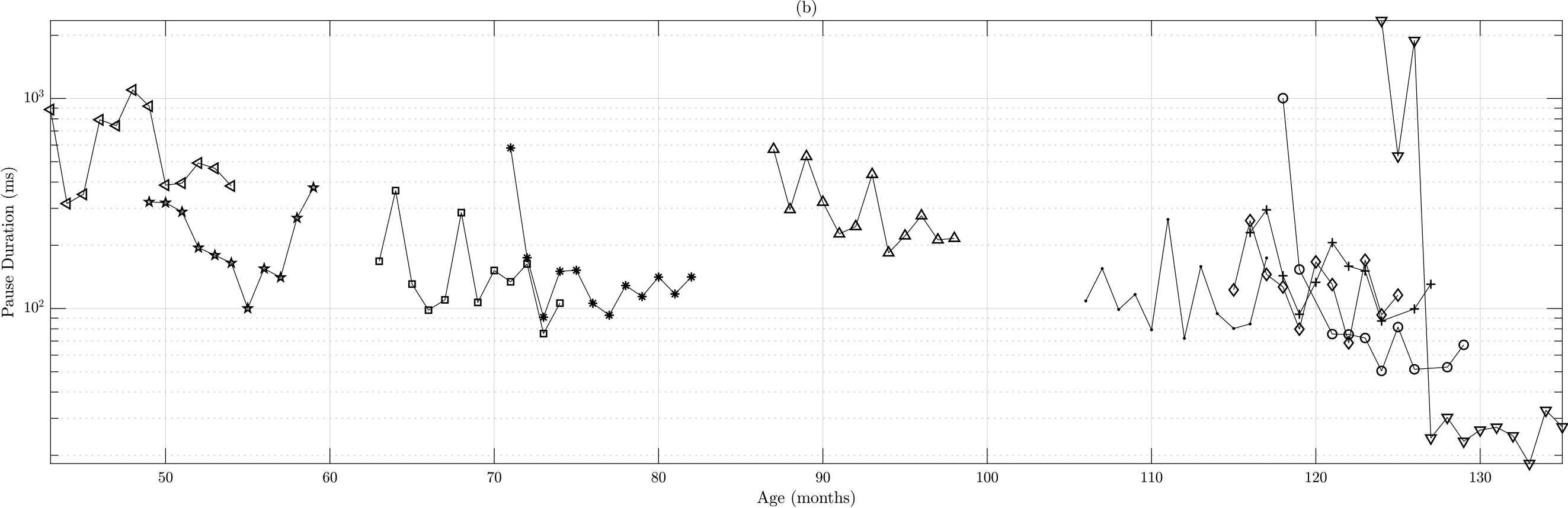}
 \caption{Monthly change of (a) mean speaking rate (including pause) and articulation rate (voiced phonemes only) and (b) mean pause duration following TBI for all subjects. The pause duration is taken directly from the inverse of the difference between the speech and articulation rate. Marker shapes unique to subject.}
 \label{fig:rates}
\end{figure*}

\begin{figure}[htbp]
 \centering
 \includegraphics[width=\linewidth]{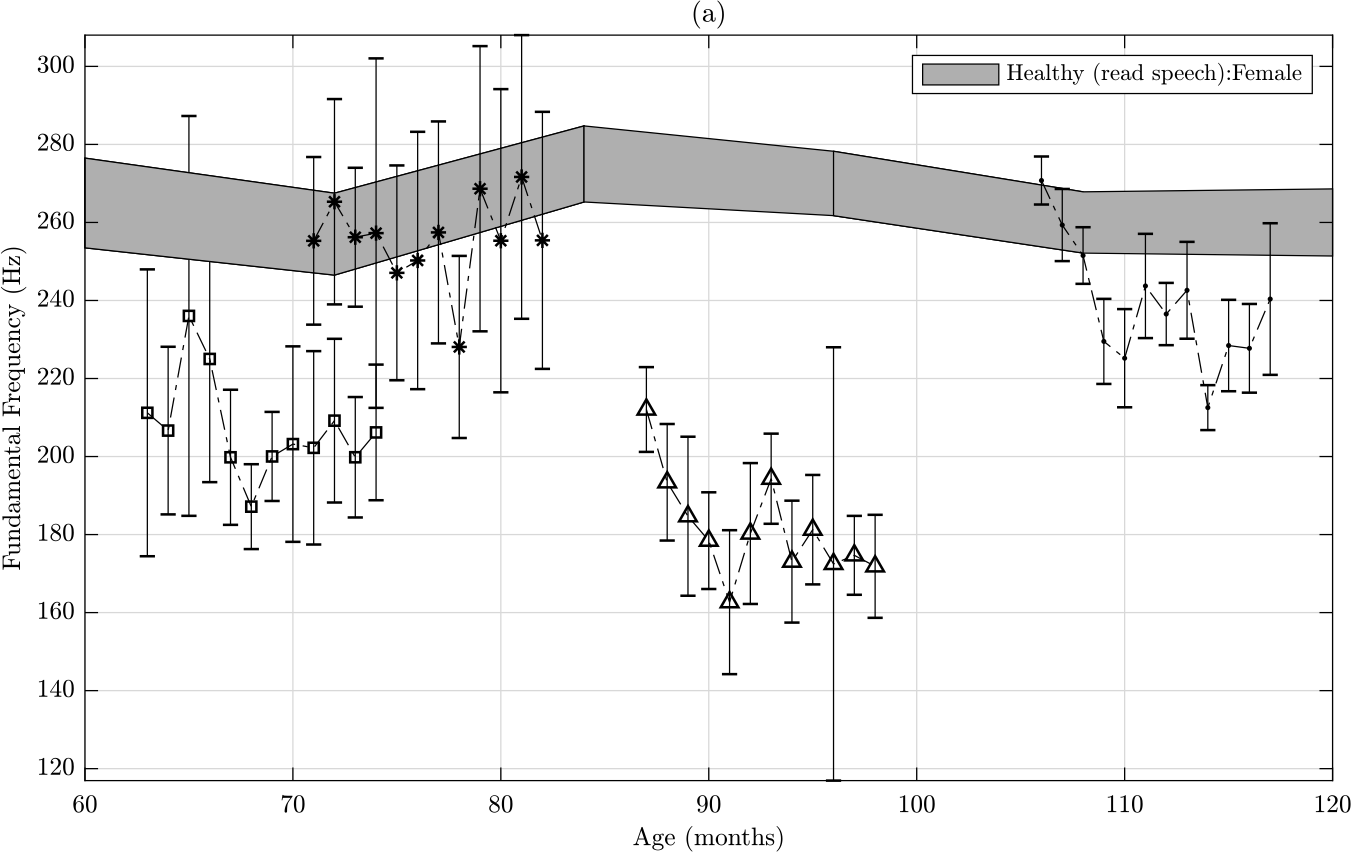}
 \includegraphics[width=\linewidth]{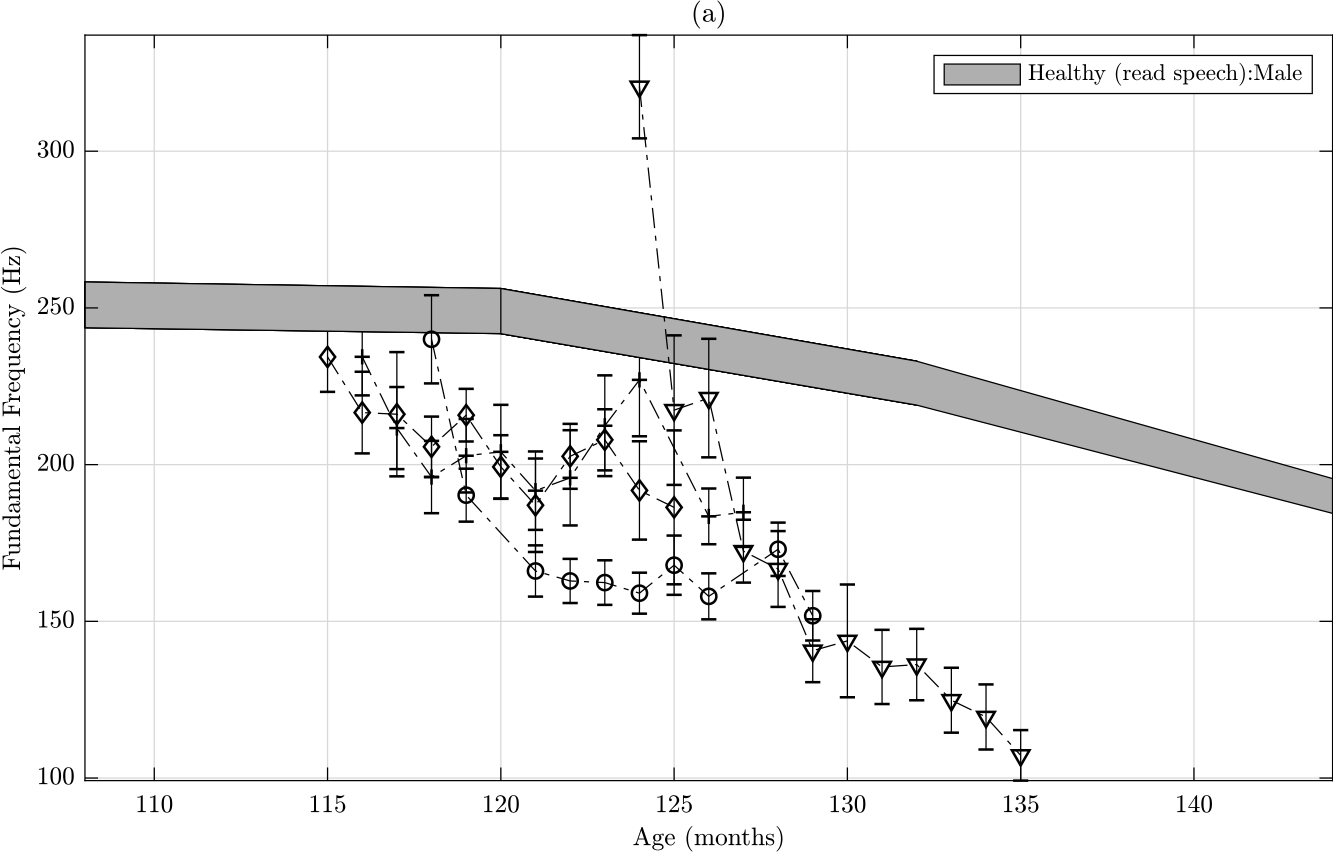}
 \caption{Monthly change of average pitch following TBI for all subjects in comparison with typical pitch during read speech. (a) female, (b) male. Marker shapes unique to subject.}
 \label{fig:f0_norms}
\end{figure}

\begin{figure}[h]
 \centering
 \includegraphics[width=\linewidth]{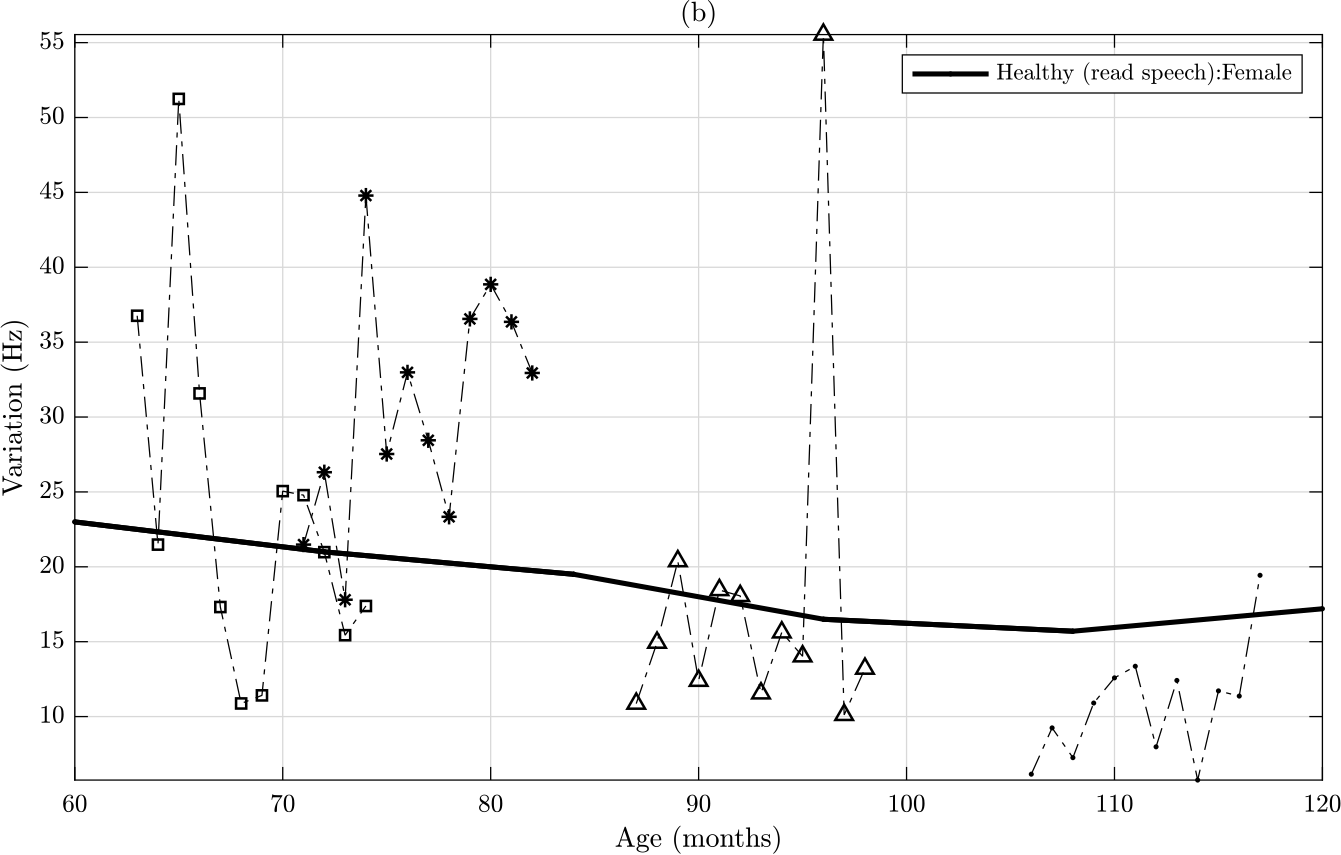}
 \includegraphics[width=\linewidth]{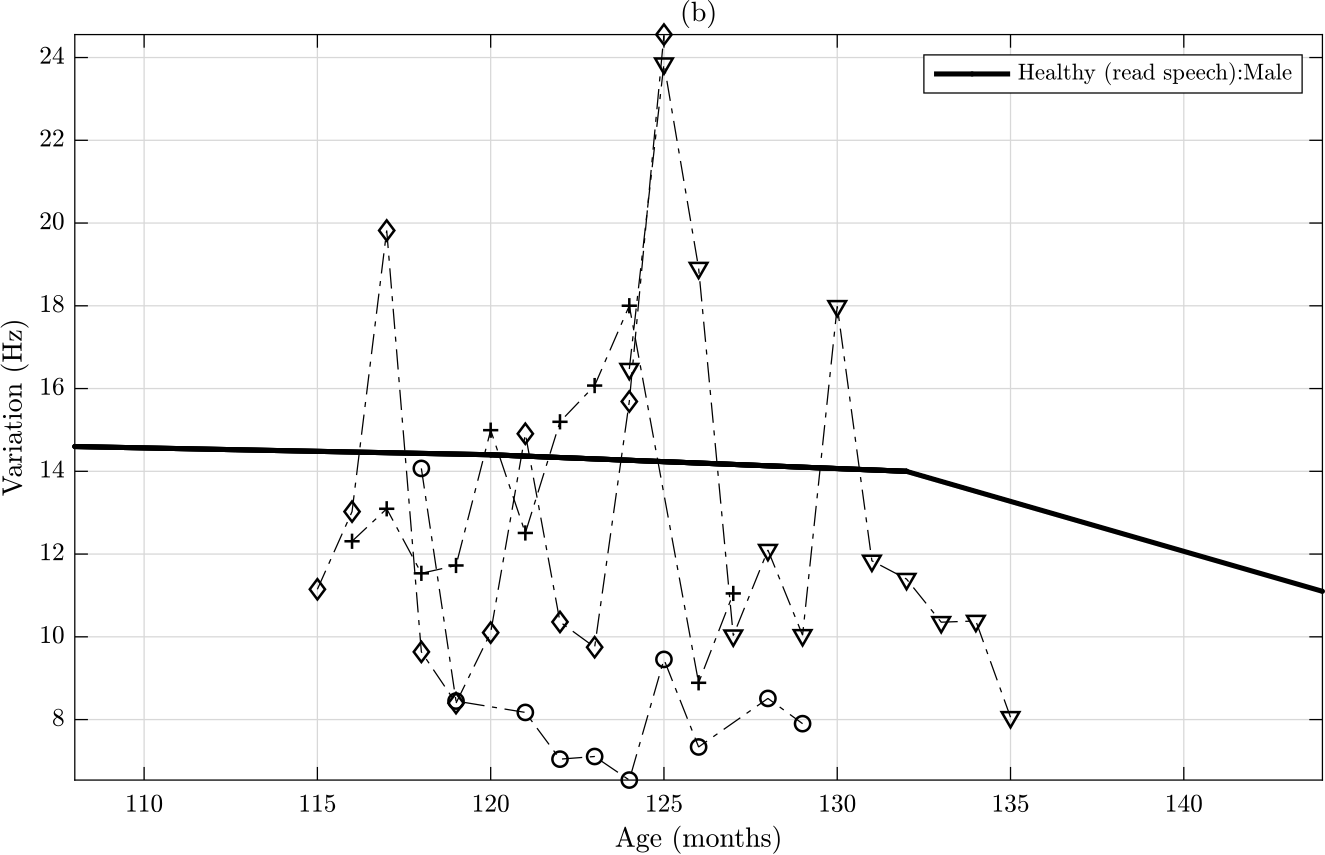}
 \caption{Monthly change in pitch variation following TBI for all subjects in comparison with typical variation during read speech. (a) female, (b) male. Marker shapes unique to subject.}
 \label{fig:f0var_norms}
\end{figure}

Table~\ref{tab:atp_group} quantitatively summarizes the trends for the cohort. The results indicate moderate, significant directional speech changes distributed across the articulatory, phonatory and prosodic subsystems. Across older children, we find statistically significant increases in speaking rate, pitch variation and diversity of phonemes spoken, while we see significant reduction in pitch and variability of consonant duration. Across younger children, we find similar reduction in variability of consonant duration, while also seeing an increase in articulatory coordination. Both younger and older children demonstrate moderate movement away from breathy and/or dysphonic voice into more typical or creaky phonation. The steadying of phone duration variability in conjunction with an increase in speech rate and a shortening of speech pause length are highly significant across the cohort. This steadying and quickening of rate is consistent with several phoneme- and syllable-based study results~~\cite{murdoch-artic-tbi,book-acoustic-dysarthria,vowels_mtbi,phone_class,Lee-Acoustics,campbell-rate}. The diversification of phoneme-use over time aligns with results shown by Campbell and Dollaghan among a pediatric and adolescent population~\cite{campbell_numwords_long}.

Via analysis of the whole cohort, we conjecture that small sample size might explain lack of numerical significance of some trends within the age-group subsets. For example, number of unique phonemes increases moderately but significantly ($p<0.001$) overall and consonant duration decreases moderately and significantly ($p=0.01$).

An empirical understanding of how speech characteristics may move in tandem with each other is shown in Fig.~\ref{fig:case_graphs}. This figure shows the change over time of phoneme diversity alongside percentage of phonemes spoken within a session being consonant. Change in values month to month are overwhelmingly synchronous in direction. If phoneme diversity increases, the portion of speech containing consonants is likely to increase as well. This suggests that both measures capture information about articulation complexity.

\subsection{Age-Dependency}

Strong correlations presented in Table~\ref{tab:atp_group} between age and several speech characteristics suggest age at injury significantly influences measurement values at the first session and may indicate the direction of change one would see in normal development. To further examine developmental influence, we compare all subjects by age and compare to previously studied speech development measurements.

A slight decrease in normalized articulation rate is not significant for either age group, indicating phoneme duration does not drastically change within the year following the TBI. However, the strong positive correlation between articulation rate and age aligns with results shown by Lee \textit{et al.}~\cite{Lee-Acoustics} that vowel and $\backslash$s$\backslash$ phoneme duration in children decrease with age. Similarly, speaking rates correlate strongly with age, but a more drastic and significant increase \textit{within} the year is exhibited by a majority of subjects. This difference indicates that pause duration decreases more quickly in the short-term following the TBI than phoneme duration. However, Nip and Green found that pause time during a narrative retell task decreased from approximately 33\% to 15\% between the age of 4 and 13 years, and that this decrease in pause time was a stronger contributor to increased speaking rate than articulation speed. The changes in pause duration over time in Fig.~\ref{fig:rates} reflect this developmental finding. Campbell \textit{et al.} hypothesized that speaking rate (and, implicitly, pause duration) may connect to `cognitive linguistic speed' and articulation rates to `articulation speed.' Therefore, the two may be dissociable among a pediatric population post-TBI~\cite{campbell-rate}. Results from the present study support that speaking rate and articulation rates do not necessarily increase in tandem, and that linguistic processing speed may be regained more rapidly than articulatory speed following TBI, as in normal development. 

Studies have shown low pitch to be a common symptom following TBI~\cite{Murdoch,book-acoustic-dysarthria}. The present results suggest that this lowered pitch continues well into the year following TBI. An atypical lowering of F0 was measured in 8 out of the 10 subjects, as compared to F0-mean and F0-standard deviation results provided by Lee \textit{et al.} (Subject 1 was out of the study's age range thus no comparison could be drawn). Many of the subjects were at or below the typical range for his/her age at the first session, and subsequently exhibited a more rapid decrease in average pitch throughout the year than what would be typical according to Lee \textit{et al.} This comparison to typical pitch change is shown in Fig.~\ref{fig:f0_norms}. One should note that this autocorrelation-based measurement of fundamental frequency may be attributed to aperiodic phonation such as glottal fry (as exhibited by the oldest subject in Fig.~\ref{fig:case_graphs}).

Correlations between time and phoneme diversity (in older children) and consonant rate variability, both highly significant when normalized, weaken when age is taken into account. Although normative data is not available for consonant duration variability, Lee \textit{et al.} find vowel duration variability decreases as a function of age. If we assume consonant duration changes similarly, we can conclude that variation decrease is somewhat expected throughout the year following TBI. However, correlations weakening with age-dependency may indicate that changes to these characteristics are more specific to recovery than to normal development.

\begin{table*}[!bthp]
\caption{Speech trend profiles of 5 subjects whose ages span the cohort's range. Measurements are grouped by speech subsystem. \\ Symbols indicate direction and strength of speech changes over one year: $\rho > 0.5$, $\rho > 0.2$ are denoted by $++$, $+$, respectively. $\rho < -0.5$, $\rho < -0.2$ are denoted by $--$, $-$, respectively. $|\rho| <0.2$ is denoted by $NT$, representing `no trend.' A $*$ denotes significant trends ($p \leq 0.05$). B $=$ below normal PCC-R threshold for age; U $=$ above; each letter represents the mean of a 4-month period.}
 \label{tab:case_symbols}
 \centering
\renewcommand{\tabcolsep}{1mm}
\resizebox{\textwidth}{!}{%
\begin{tabular}{@{}llll|l|lll|llll|llll@{}}
\toprule
\textbf{} & \textbf{} & \textbf{} & \textbf{} & \textbf{} & \textbf{Phonation} & \textbf{} & \textbf{} & \textbf{Prosody} & \textbf{} & \textbf{} & \textbf{} & \textbf{Articulation} & \textbf{} & \textbf{} & \textbf{} \\ \midrule
\textbf{Subject} & \textbf{Sex} & \textbf{Age-Inj.} & \textbf{Age-S1.} & \textbf{measure} & \textbf{CPP} & \textbf{H2-H1} & \textbf{F0} & \textbf{VW Dur.} & \textbf{CO Dur.} & \textbf{Art. Rate} & \textbf{Sp. Rate} & \textbf{F1-F3 Corr.} & \textbf{\# Unq. Ph.} & \textbf{CO \%} & \textbf{PCC-R} \\ \midrule
1 & M & 21 & 43 & mean: & + + & - & - & - - & \textit{\textbf{- -*}} & \textit{\textbf{+ +*}} & + & + + & + & \textit{\textbf{+ +*}} & B B B \\
 & & & & var: & \textit{\textbf{+ +*}} & - & - - & - & \textit{\textbf{- -*}} & & & & & & \\ \midrule
3 & F & 63 & 63 & mean: & \textit{\textbf{- -*}} & - & - & - & - & + & + & NT & + & NT & B B U \\
 & & & & var: & \textit{\textbf{- -*}} & - & - & - & - & & & & & & \\ \midrule
5 & F & 85 & 87 & mean: & \textit{\textbf{- -*}} & + & \textit{\textbf{- -*}} & \textit{\textbf{- -*}} & \textit{\textbf{- -*}} & \textit{\textbf{+ +*}} & \textit{\textbf{+ +*}} & NT & \textit{\textbf{+ +*}} & + + & U U U \\
 & & & & var: & - & NT & + & - - & - - & & & & & & \\ \midrule
6 & F & 102 & 106 & mean: & \textit{\textbf{+ +*}} & - & \textit{\textbf{- -*}} & \textit{\textbf{+ +*}} & \textit{\textbf{- -*}} & \textit{\textbf{- -*}} & - - & - & - & NT & U U U \\
 & & & & var: & + + & NT & + + & \textit{\textbf{+ +*}} & NT & & & & & & \\ \midrule
10 & M & 123 & 124 & mean: & \textit{\textbf{+ +*}} & \textit{\textbf{+ +*}} & \textit{\textbf{- -*}} & - & - & + & \textit{\textbf{+ +*}} & + & \textit{\textbf{+ +*}} & - & U U U \\
 & & & & var: & \textit{\textbf{+ +*}} & \textit{\textbf{+ +*}} & \textit{\textbf{+ +*}} & - - & \textit{\textbf{- -*}} & & & & & & \\ \bottomrule
\end{tabular}
}
\end{table*}

\subsection{Case Studies: Within-Subject Trends}

A comprehensive feature set is necessary to analyze the different speech production mechanisms and how they may interact over time. Subsystem interactions are discussed case-by-case for 5 subjects whose ages span the cohort's range. 

\begin{figure}[htbp]
 \centering
 \includegraphics[width=\linewidth]{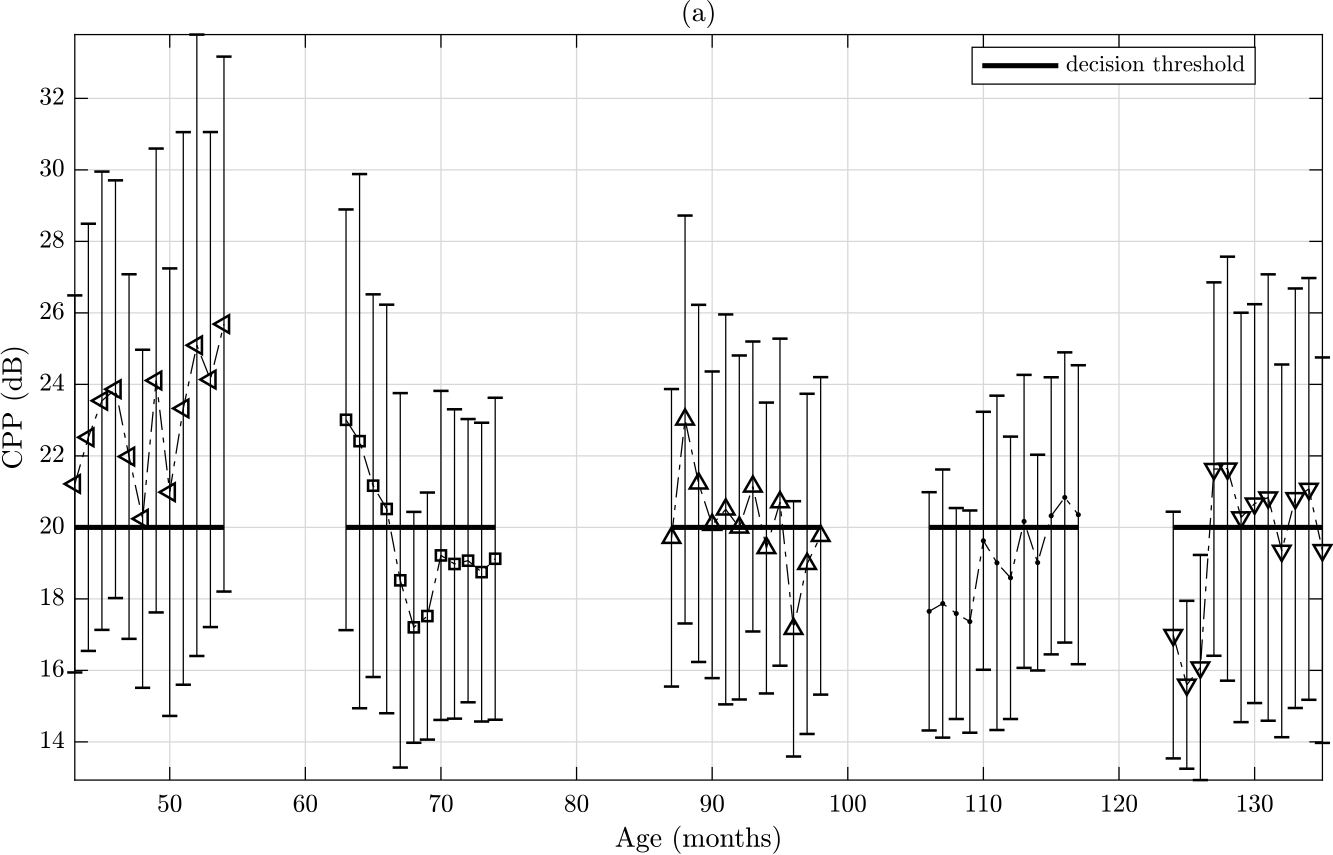}
 \includegraphics[width=\linewidth]{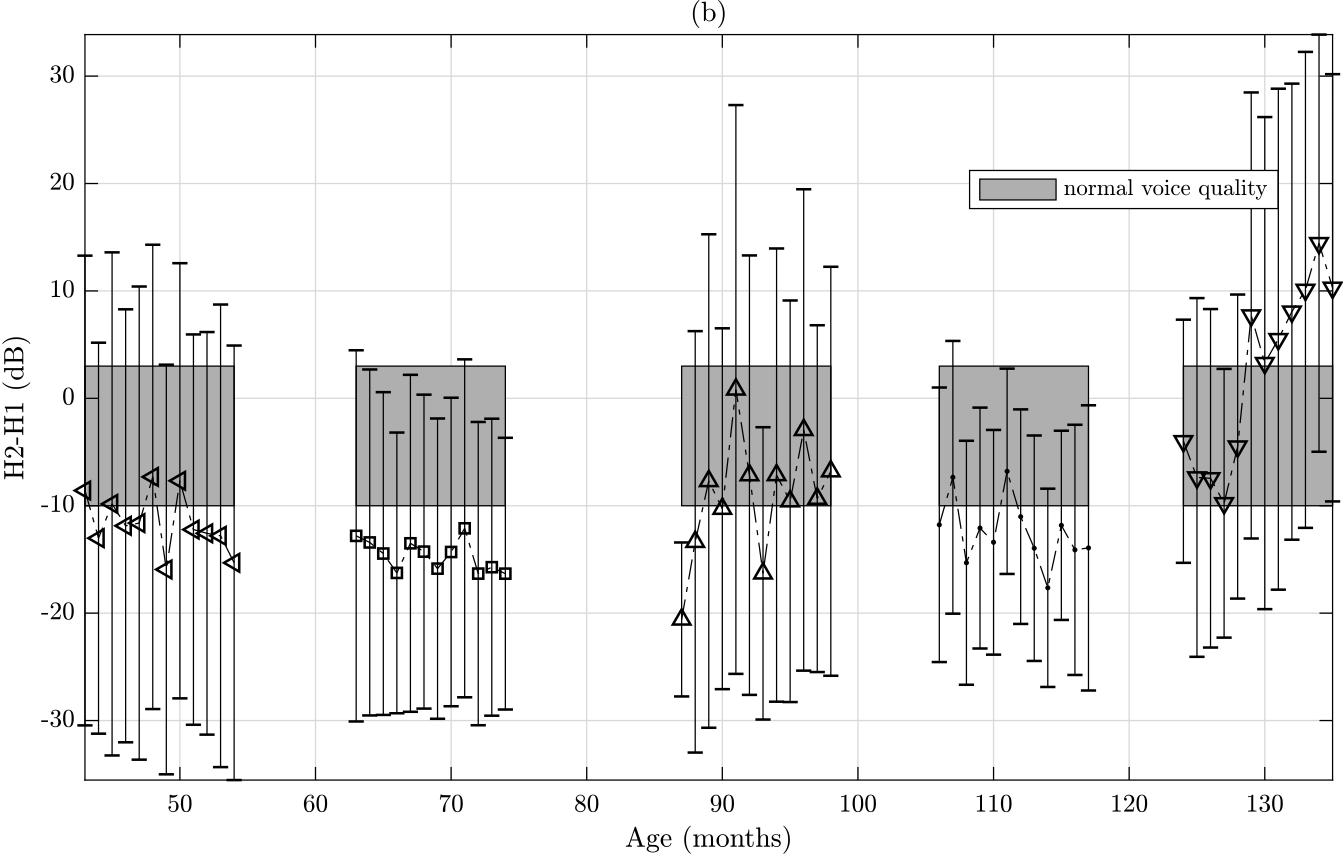}
 \includegraphics[width=\linewidth]{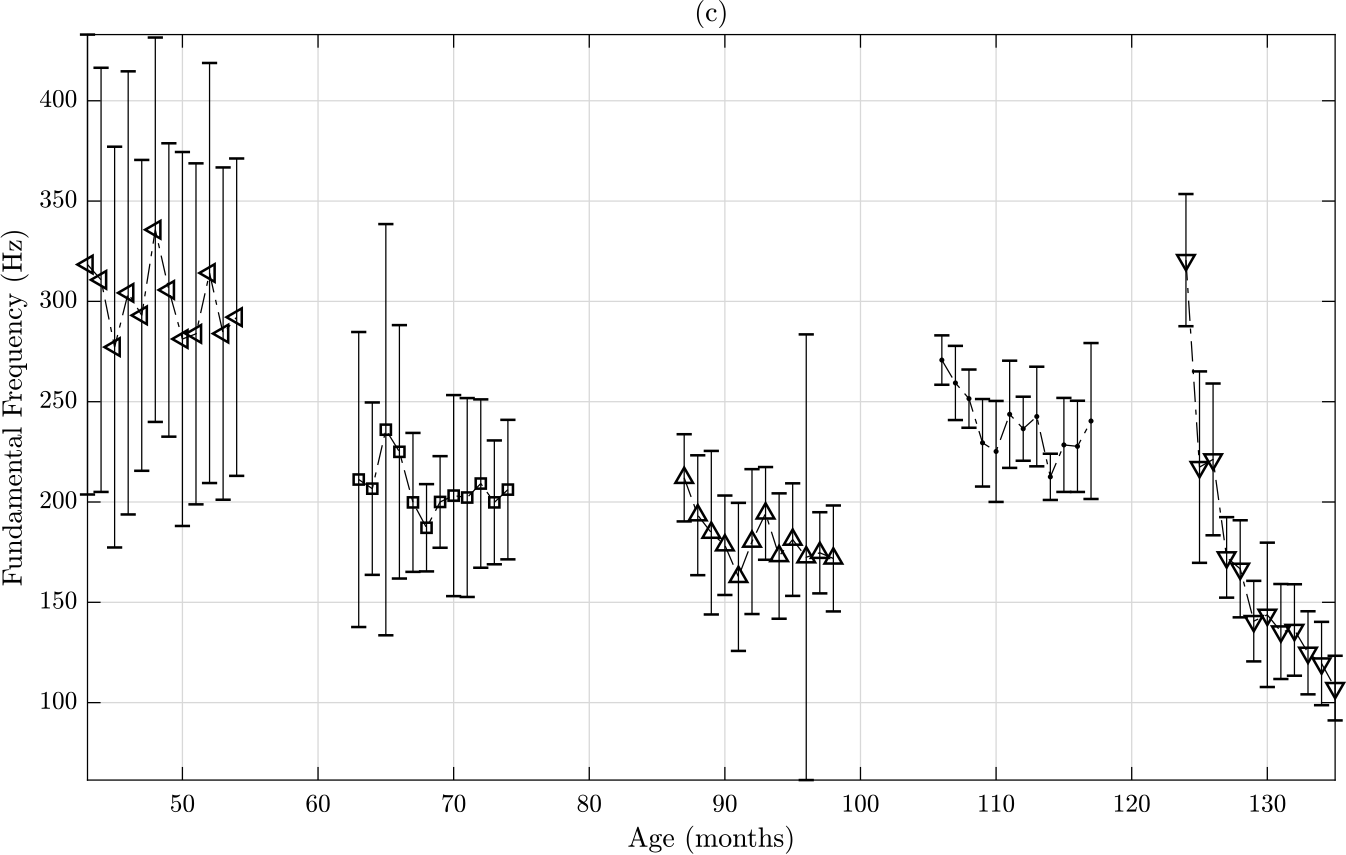}
 \includegraphics[width=\linewidth]{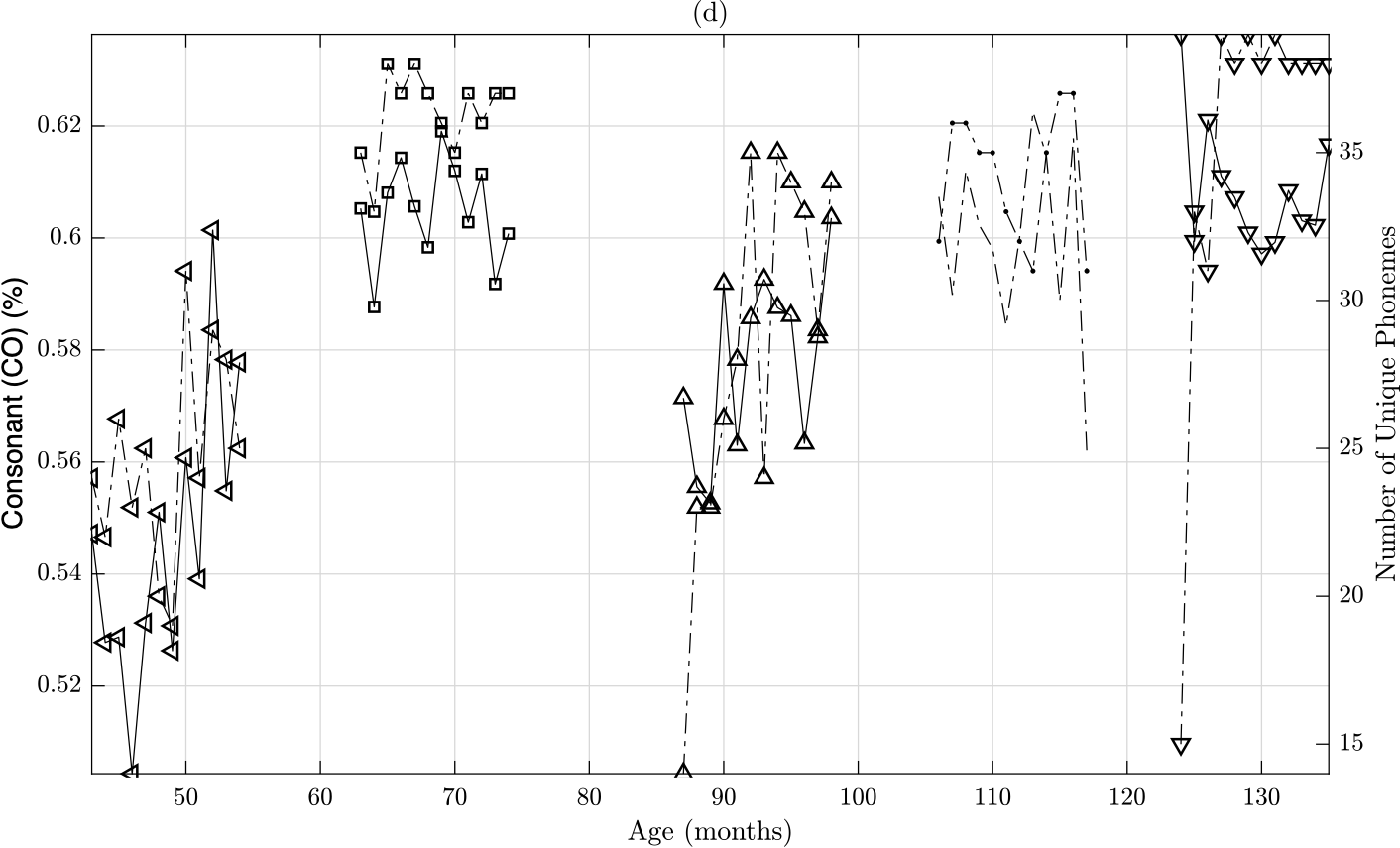}
 \caption{Monthly change of five phonation measures. Mean and standard deviation (error bars) of each session are shown for 5 subjects whose ages span the cohort's range. (a) Cepstral Peak Prominence (CPP), (b) H2-H1 Amplitude, (c) Pitch (F0), (d) Percentage of phonemes uttered classified as consonants (CO) (left axis, solid line) and number of unique phonemes uttered (right axis, dashed line). Marker shapes unique to subject.}
 \label{fig:case_graphs}
\end{figure}

Table~\ref{tab:case_symbols} symbolically represents the comprehensive set of speech trends calculated for these 5 subjects via the correlation procedure described in Section \ref{method_sum}. Positive trends (i.e.. an increase in value over time) are succinctly represented by the $+$ sign while negative trends (i.e. a decrease in value over time) are represented by a $-$ sign. A numerical version of this table for all ten subjects is presented in Table~\ref{tab:appendix}. Fig.~\ref{fig:case_graphs} shows monthly change over time of characteristics within the phonation and articulation subsystems. 

\subsubsection{Subject 1} This child, the youngest in the cohort, experienced a TBI at 21 months old and was able to start speech sampling sessions at 43 months. Over the following year, he increases his speaking rate more so than articulation rate, indicating an increase in articulation speed over time as suggested by Campbell \textit{et al.}~\cite{campbell-rate}. His consonant rate steadies over time, suggesting an increase in timing control~\cite{book-acoustic-dysarthria, McCauley-RecsTBI}. The moderate increase in complexity of formant interaction is consistent with moderate increases in the ratio of consonants to vowels and number of unique phonemes produced per session, indicating a trend toward more complex articulation ability coupled with an improvement in linguistic processing ability. The child was reported to consistently be under the PCC-R normative curve for her age throughout this period~\cite{PCC-R}. Although details describing distance from this curve are not available, the acoustic articulatory and prosodic measurements suggest that perceptual intelligibility is changing over time despite PCC-R values remaining below normal. These subsystems appear to be more affected than the phonatory subsystem; this child exhibited healthy phonation measurements throughout the speech sampling period.

\subsubsection{Subject 3} This child experienced a TBI as well as started speech sampling sessions at 63 months old. Following the TBI, CPP measurements indicate movement toward a more breathy or dysphonic voice over time, but glottal constriction and pitch levels do not change. A slight decrease in pitch variation occurs, normal for this subject's age group. The articulatory and prosodic subsystems appear to be less altered, with slight but non-significant improvements in overall speaking and articulation rates and phoneme diversity. These slight improvements may still be indicative of improvement in motor control, as the PCC-R measure does cross the normative threshold for her age.

\subsubsection{Subject 5} This child experienced a TBI at 85 months old and started speech sampling sessions 2 months later. Over the course of the first year of recovery, she exhibits significant change across all subsystems. A slight but significant increase in dysphonic or breathy voice quality in conjunction with a decrease in pitch occurs. The child's starting pitch is much lower than typical for her age but pitch variation remains typical. The subject's PCC-R sits above the normative threshold from the first session, and no noticeable change in F1-F3 complexity is exhibited. However, phonemes diversity and consonant consonants significantly increases over time, indicating some increased articulatory coordination abilities. Articulatory speed increases for both vowels and consonants, and intraphrase pause time decreases. Therefore, increases in both cognitive and motor processing speed are suggested.

\subsubsection{Subject 6} This child experienced a TBI at 102 months old and began speech sampling sessions 4 months later. Throughout the year, this child exhibits prosodic change in the opposite direction of the expected development trend. Articulation rates (due to increase in both vowel and consonant duration) decrease significantly, while pause duration decreases less so (see. Fig.~\ref{fig:rates}). Pitch levels and variation are near typical range. An increase average phonation periodicity reaches levels above the healthy threshold. This child does not exhibit significant change in articulation complexity by any acoustic measure, and PCC-R scores are reported above the normative threshold throughout the sampling sessions.   

\subsubsection{Subject 10} The oldest child in the cohort was 123 months in age at injury and began sampling sessions 1 month later. He exhibits large changes within all three subsystems during the first 3 months of recovery. A drastic shortening of pause length occurs, suggesting an improvement in linguistic processing speed. Phoneme duration variation steadies significantly while phoneme diversity increases by 19 phonemes. These same 3 months see a drastic drop in average pitch and consistent breathy phonation. The subsequent nine months begin with a jump in glottal-constricted creaky voice paired with more normal levels of breathiness. The continued lowering of pitch may be due to hormonal changes. Formant interaction complexity and consonant occurrence did not change significantly, and PCC-R values remained above the age-normative threshold.

These case studies highlight the varied recovery trajectories and possible variants of dysarthric and dysphonic speech a child may obtain following TBI. Comparison with the cohort-wide and age-dependent trends discussed above allow these subsystem measurements to be compared and understood within the necessary contexts. Comparison between the acoustic features and the subsystems they represent demonstrate the advantage and necessity of sensitive, objective and therefore quickly and incrementally repeatable measurements to supply information far beyond what perceptual measurements like PCC-R alone can provide.

\section{Future Work}

Several features that may correlate significantly with recovery also correlated significantly with age. Thus, future work expanding upon age-specific normative acoustic and linguistic measurements (as in~\cite{McCauley-RecsTBI,PCC-R,Lee-Acoustics, Coplan-Unclear}.) or conducting longitudinal case-control studies (as in~\cite{campbell-rate}) would aid in judging the efficacy of age-specific, instrumentally-advised trajectories. For example, formant-track correlation complexity increases in the younger children, whereas the older children are able to produce a more diverse set of sounds after a year. Both indicate an increase in motor coordination abilities but are different manifestations, perhaps influenced by other developmental factors. Normative measurements would help determine the perceptual relevance and strength of a quantitative trend. 

It is important to note the shortcomings of this particular method applied to this specific dataset. The trends exhibited by this small cohort may not generalize to a larger population and are likely not suitable stand-alone predictors of the wide-ranging symptoms and recovery patterns children exhibit following a TBI. Additionally, this dataset is unique in that it provided an opportunity to apply acoustic features longitudinally, but it was not originally designed for such acoustic analysis. Therefore, the feature set was constrained to work with spontaneous child speech of variable length; it is by no means an exhaustive set~\cite{book-acoustic-dysarthria}. The recordings were also subject to a multitude of environmental circumstances that may have affected the discourse and speaking style of a child during a sampling session, possibly confounding the acoustics of the vocal signals. Finally, the utilized diarization and segmentation tools used are trained on healthy adult speech, and hence likely to degrade in performance on child or disordered speech. Although performance may be consistent within a subject, research into how acoustic analysis tools perform on atypical speech would allow for more robust population analysis and help create more reliable and inclusive assessment systems. 

\section{Conclusion}
This study aimed to track the acoustic changes within and across speech subsystems of children who had suffered a severe TBI. The results support existing evidence that the brain's widely distributed speech network is often impacted following a TBI, and features measuring the nuanced, outward manifestations of linguistic processing and motor control may be viable indicators of trauma to the brain.  

Specifically, longitudinal, repeated measurement of acoustic speech features within the articulatory, prosodic and phonatory subsystems provided nuanced information about how the control of these subsystems may change or improve over time. The application of these features to individuals demonstrated their ability to comprehensively capture the variant speech impairments and recovery trajectories possible following TBI. Cohort-wide trends demonstrated that variations of impairment within the articulatory and prosodic subsystems are simultaneously possible, and that linguistic processing, articulatory coordination, and articulatory speed do not necessarily improve in tandem. Changes over time within the phonatory subsystem were significant across the cohort and appeared to be independent from changes to speech motor-control. Age-dependent characteristics of speech appeared to guide recovery trajectories in many cases, signaling that continued development in the normative direction can still occur following pediatric TBI. However, baseline measurements and varying trajectory rates out of typical range furthers evidence that a TBI obtained during childhood may impact subsequent speech development. This study thus provides a baseline approach for longitudinal acoustic analysis of speech and calls for a prospective study on the use of acoustic features in pediatric speech tracking following traumatic brain injury.

\goodbreak
\section*{Acknowledgements}

The authors thank Drs. Michael Brandstein, Bengt Borgstrom, Malcolm Slaney and Jonathan Berger for their contribution to and feedback on this work.

\bibliographystyle{IEEEtran}

\bibliography{mybib}

\newpage

\begin{sidewaystable*}
\centering
\caption{Normalized Pearson correlation measures with time for all subjects in the cohort. \\ $\rho$ = correlation size, $p$ = significance. Skew is only calculated for H2-H1 measurement.}
\label{tab:appendix}
\resizebox{\textheight}{!}{%
\begin{tabular}{@{}lllllllllllllll@{}}
\toprule
 & \textbf{} & \textbf{} & \textbf{} & \textbf{} & \textbf{Phonation} & \textbf{} & \textbf{} & \textbf{Prosody} & \textbf{} & \textbf{} & \textbf{} & \textbf{Articulation} & \textbf{} \\ \midrule
\multicolumn{1}{l|}{\textbf{Subject}} & Sex & Age-Inj. (m(y)) & \multicolumn{1}{l|}{Age-S1. (m(y))} & \multicolumn{1}{l|}{\begin{tabular}[c]{@{}l@{}}Measure:\\ $\rho$ ($p$)\end{tabular}} & CPP & H2-H1 & \multicolumn{1}{l|}{F0} & VW Dur. & CO Dur. & Art. Rate & \multicolumn{1}{l|}{Sp. Rate} & \# Unq Ph & CO \% & F1-F3 Corr.\\ \midrule
\multicolumn{1}{l|}{\textbf{1}} & M & 21 (1.8) & \multicolumn{1}{l|}{43 (3.6)} & \multicolumn{1}{l|}{Mean:} & 0.508 (0.092) & -0.434 (0.158) & \multicolumn{1}{l|}{-0.383 (0.219)} & -0.529 (0.07) & -0.644 (0.024) & 0.656 0.020) & \multicolumn{1}{l|}{0.338 (0.282)} & 0.386 (0.215) & 0.644 (0.024) & -0.501\\
\multicolumn{1}{l|}{\textbf{}} & & & \multicolumn{1}{l|}{} & \multicolumn{1}{l|}{Variance:} & 0.740 (0.006) & -0.372 (0.234) & \multicolumn{1}{l|}{-0.521 0.08)} & -0.266 (0.4-3) & -0.602 (0.038) & & \multicolumn{1}{l|}{} & & \\
\multicolumn{1}{l|}{\textbf{}} & & & \multicolumn{1}{l|}{} & \multicolumn{1}{l|}{Skew:} & & 0.364 (0.245) & \multicolumn{1}{l|}{} & & & & \multicolumn{1}{l|}{} & & \\ \midrule
\multicolumn{1}{l|}{\textbf{2}} & M & 47 (3.9) & \multicolumn{1}{l|}{48 (4.0)} & \multicolumn{1}{l|}{} & 0.701 (0.016) & -0.603 (0.050) & \multicolumn{1}{l|}{-0.743 (0.009)} & 0.446 (0.169) & -0.206 (0.543) & -0.141 (0.679) & \multicolumn{1}{l|}{0.190 (0.576)} & -0.2744 (0.414) & 0.070 (0.837) & -0.325\\
\multicolumn{1}{l|}{\textbf{}} & & & \multicolumn{1}{l|}{} & \multicolumn{1}{l|}{} & 0.501 (0.117) & -0.515 (0.105) & \multicolumn{1}{l|}{-0.339 (0.349)} & 0.044 (0.897) & -0.618 (0.043) & & \multicolumn{1}{l|}{} & & \\
\multicolumn{1}{l|}{\textbf{}} & & & \multicolumn{1}{l|}{} & \multicolumn{1}{l|}{} & & 0.581 (0.061) & \multicolumn{1}{l|}{} & & & & \multicolumn{1}{l|}{} & & \\ \midrule
\multicolumn{1}{l|}{\textbf{3}} & F & 63 (5.3) & \multicolumn{1}{l|}{63 (5.3)} & \multicolumn{1}{l|}{} & -0.679 (0.015) & -0.368 (0.239) & \multicolumn{1}{l|}{-0.456 (0.137)} & -0.471 (0.122) & -0.293 (0.355) & 0.261 (0.412) & \multicolumn{1}{l|}{0.450 (0.143)} & 0.299 (0.346) & -0.009 (0.978) & -0.200\\
\multicolumn{1}{l|}{\textbf{}} & & & \multicolumn{1}{l|}{} & \multicolumn{1}{l|}{} & -0.668 (0.017) & -0.484 (0.111) & \multicolumn{1}{l|}{-0.366 (0.242)} & -0.416 (0.179) & -0.369 (0.239) & & \multicolumn{1}{l|}{} & & \\
\multicolumn{1}{l|}{\textbf{}} & & & \multicolumn{1}{l|}{} & \multicolumn{1}{l|}{} & & 0.624 (0.030) & \multicolumn{1}{l|}{} & & & & \multicolumn{1}{l|}{} & & \\ \midrule
\multicolumn{1}{l|}{\textbf{4}} & F & 70 (5.8) & \multicolumn{1}{l|}{71 (5.9)} & \multicolumn{1}{l|}{} & 0.312 (0.324) & -0.303 (0.339) & \multicolumn{1}{l|}{0.243 (0.447)} & -0.116 (0.719) & 0.106 (0.743) & -0.047 (0.884) & \multicolumn{1}{l|}{0.318 (0.314)} & 0.472 (0.121) & 0.448 (0.144) & -0.192\\
\multicolumn{1}{l|}{\textbf{}} & & & \multicolumn{1}{l|}{} & \multicolumn{1}{l|}{} & 0.867 (0.000) & 0.013 (0.969) & \multicolumn{1}{l|}{0.067 (0.836)} & -0.578 (0.049) & -0.498 (0.100) & & \multicolumn{1}{l|}{} & & \\
\multicolumn{1}{l|}{\textbf{}} & & & \multicolumn{1}{l|}{} & \multicolumn{1}{l|}{} & & 0.153 (0.636) & \multicolumn{1}{l|}{} & & & & \multicolumn{1}{l|}{} & & \\ \midrule
\multicolumn{1}{l|}{\textbf{5}} & F & 85 (7.1) & \multicolumn{1}{l|}{87 (7.3)} & \multicolumn{1}{l|}{} & -0.589 (0.044) & 0.362 (0.247) & \multicolumn{1}{l|}{-0.563 (0.056)} & -0.747 (0.005) & -0.677 (0.016) & 0.831 (0.001) & \multicolumn{1}{l|}{0.765 (0.004)} & 0.767 (0.004) & 0.557 (0.060) & -0.067\\
\multicolumn{1}{l|}{\textbf{}} & & & \multicolumn{1}{l|}{} & \multicolumn{1}{l|}{} & -0.452 (0.141) & 0.163 (0.612) & \multicolumn{1}{l|}{0.466 (0.127)} & -0.551 (0.063) & -0.551 (0.063) & & \multicolumn{1}{l|}{} & & \\
\multicolumn{1}{l|}{\textbf{}} & & & \multicolumn{1}{l|}{} & \multicolumn{1}{l|}{} & & -0.410 (0.187) & \multicolumn{1}{l|}{} & & & & \multicolumn{1}{l|}{} & & \\ \midrule
\multicolumn{1}{l|}{\textbf{6}} & F & 102 (8.5) & \multicolumn{1}{l|}{106 (8.8)} & \multicolumn{1}{l|}{} & 0.872 (0.000) & -0.387 (0.218) & \multicolumn{1}{l|}{-0.615 ( 0.033)} & 0.726 (0.007) & -0.644 (0.024) & -0.707 (0.010) & \multicolumn{1}{l|}{-0.55 (0.864)} & -0.045 (0.890) & -0.192 (0.550) & 0.055\\
\multicolumn{1}{l|}{\textbf{}} & & & \multicolumn{1}{l|}{} & \multicolumn{1}{l|}{} & 0.511 (0.090) & -0.100 (0.756) & \multicolumn{1}{l|}{0.529 (0.077)} & 0.589 (0.043) & 0.102 (0.752) & & \multicolumn{1}{l|}{} & & \\
\multicolumn{1}{l|}{\textbf{}} & & & \multicolumn{1}{l|}{} & \multicolumn{1}{l|}{} & & 0.498 (0.099) & \multicolumn{1}{l|}{} & & & & \multicolumn{1}{l|}{} & & \\ \midrule
\multicolumn{1}{l|}{\textbf{7}} & M & 114 (9.5) & \multicolumn{1}{l|}{115 (9.6)} & \multicolumn{1}{l|}{} & -0.103 (0.762) & 0.500 (0.117) & \multicolumn{1}{l|}{-0.659 (0.027)} & -0.341 (0.304) & 0.202 (0.552) & 0.144 (0.673) & \multicolumn{1}{l|}{0.343 (0.301)} & 0.501 (0.116) & 0.354 (0.285) & -0.178\\
\multicolumn{1}{l|}{\textbf{}} & & & \multicolumn{1}{l|}{} & \multicolumn{1}{l|}{} & 0.389 (0.238) & 0.253 (0.452) & \multicolumn{1}{l|}{0.415 (0.204)} & -0.415 (0.204) & -0.250 (0.046) & & \multicolumn{1}{l|}{} & & \\
\multicolumn{1}{l|}{\textbf{}} & & & \multicolumn{1}{l|}{} & \multicolumn{1}{l|}{} & & -0.355 (0.283) & \multicolumn{1}{l|}{} & & & & \multicolumn{1}{l|}{} & & \\ \midrule
\multicolumn{1}{l|}{\textbf{8}} & M & 115 (9.6) & \multicolumn{1}{l|}{116 (9.7)} & \multicolumn{1}{l|}{} & 0.051 (0.881) & 0.71 (0.014) & \multicolumn{1}{l|}{-0.517 (0.103)} & -0.225 (0.505) & -0.448 (0.167) & 0.369 (0.263) & \multicolumn{1}{l|}{0.658 (0.028)} & 0.237 (0.483) & 0.226 (0.503) & -0.405\\
\multicolumn{1}{l|}{\textbf{}} & & & \multicolumn{1}{l|}{} & \multicolumn{1}{l|}{} & -0.281 (0.403) & 0.621 (0.041) & \multicolumn{1}{l|}{0.403 (0.219)} & -0.286 (0.394) & -0.535 (0.090) & & \multicolumn{1}{l|}{} & & \\
\multicolumn{1}{l|}{\textbf{}} & & & \multicolumn{1}{l|}{} & \multicolumn{1}{l|}{} & & -0.304 (0.363) & \multicolumn{1}{l|}{} & & & & \multicolumn{1}{l|}{} & & \\ \midrule
\multicolumn{1}{l|}{\textbf{9}} & M & 116 (9.7) & \multicolumn{1}{l|}{118 (9.8)} & \multicolumn{1}{l|}{} & 0.501 (0.135) & 0.814 (0.004) & \multicolumn{1}{l|}{-0.644 (0.044)} & -0.419 (0.229) & -0.296 (0.407) & 0.368 (0.296) & \multicolumn{1}{l|}{0.690 (0.027)} & 0.552 (0.097) & 0.610 (0.061) & -0.103\\
\multicolumn{1}{l|}{\textbf{}} & & & \multicolumn{1}{l|}{} & \multicolumn{1}{l|}{} & 0.664 (0.036) & 0.775 (0.009) & \multicolumn{1}{l|}{0.857 (0.002)} & -0.381 (0.277) & -0.481 (0.159) & & \multicolumn{1}{l|}{} & & \\
\multicolumn{1}{l|}{\textbf{}} & & & \multicolumn{1}{l|}{} & \multicolumn{1}{l|}{} & & -0.743 (0.013) & \multicolumn{1}{l|}{} & & & & \multicolumn{1}{l|}{} & & \\ \midrule
\multicolumn{1}{l|}{\textbf{10}} & M & 123 (10.3) & \multicolumn{1}{l|}{124 (10.3)} & \multicolumn{1}{l|}{} & 0.580 (0.048) & 0.894 (0.000) & \multicolumn{1}{l|}{-0.850 (0.000)} & -0.333 (0.291) & -0.214 (0.504) & 0.244 (0.445) & \multicolumn{1}{l|}{0.696 (0.012)} & 0.630 (0.028) & -0.395 (0.204) & -0.295\\
\multicolumn{1}{l|}{\textbf{}} & & & \multicolumn{1}{l|}{} & \multicolumn{1}{l|}{} & 0.720 (0.009) & 0.598 (0.040) & \multicolumn{1}{l|}{0.594 (0.041)} & -0.565 (0.056) & -0.634 (0.025) & & \multicolumn{1}{l|}{} & & \\
\textbf{} & & & & & & -0.695 (0.012) & & & & & & & \\ \bottomrule
\end{tabular}%
}
\end{sidewaystable*}

\end{document}